\documentclass[conference]{IEEEtran}
\IEEEoverridecommandlockouts
\usepackage{cite}
\usepackage{amsmath,amssymb,amsfonts}
\usepackage{algorithmic}
\usepackage{graphicx}
\usepackage{textcomp}
\usepackage{xcolor}
\usepackage{cite}
\usepackage{amssymb}
\usepackage{amsfonts}
\usepackage{amsmath}
\usepackage{amsthm}
\usepackage{mathrsfs}
\usepackage{bm}
\usepackage{verbatim}
\usepackage{autobreak} 
\usepackage{graphicx} 
\usepackage{float} 
\usepackage{subfigure} 
\usepackage{tabularx}
\def\BibTeX{{\rm B\kern-.05em{\sc i\kern-.025em b}\kern-.08em
    T\kern-.1667em\lower.7ex\hbox{E}\kern-.125emX}}
\begin{document}

\title{Demo: Reconfigurable Distributed Antennas and Reflecting Surface (RDARS)-aided Integrated Sensing and Communication System\\
}
\author{Jintao Wang, Chengwang Ji, Jiajia Guo, Shaodan Ma \\
State Key Laboratory of Internet of Things for Smart City (SKL-IOTSC)\\
Department of Electrical and Computer Engineering, University of Macau, Macao SAR, China\\
Emails: \{wang.jintao,~ji.chengwang\}@connect.um.edu.mo, \{jiajiaguo,~shaodanma\}@um.edu.mo
}

\maketitle

\IEEEpubid{\begin{minipage}[t]{\textwidth}\ \\[12pt] \centering
  \copyright \ 2023 IEEE. Personal use of this material is permitted. Permission from IEEE must be obtained for all other uses, in any current or future media, including reprinting/republishing this material for advertising or promotional purposes, creating new collective works, for resale or redistribution to servers or lists, or reuse of any copyrighted component of this work in other works.
\end{minipage}} 
\begin{abstract}

Integrated sensing and communication (ISAC) system  has been envisioned as a promising technology to be applied in future 
applications requiring both communication and high-accuracy
sensing. Different from most research focusing on theoretical analysis and optimization in the area of ISAC, we implement a reconfigurable distributed antennas and reflecting surfaces (RDARS)-aided ISAC system prototype to achieve the dual-functionalities with the communication signal. A RDARS, composed of programmable elements capable of switching between reflection mode and connected mode, is introduced to assist in uplink signal transmission and sensing. 
The developed RDARS-aided ISAC prototype achieves reliable user localization without compromising the communication rate, showcasing its potential for future 6G systems.
\end{abstract}


\section{Background}
With the explosive growth in demands of wireless communication, the integrated sensing and communications (ISAC) system has been envisioned as one of the key enabling technologies in 6G systems.
ISAC leverages a shared platform to enhance spectrum and hardware efficiencies by integrating communication and sensing functionalities, effectively utilizing shared resources.
 To achieve simultaneous sensing and communications, most of the current research focuses on the theoretical study of ISAC systems, such as joint waveform design, joint signal processing, and performance optimization \cite{liu2022integrated}.    
 In this paper, we design a novel reconfigurable surface-aided integrated sensing and communication prototype to simultaneously achieve communication and user localization functions. 
 Different from the conventional reconfigurable intelligent surface (RIS)-aided ISAC systems\cite{Shao2022localization}, our prototype utilizes the recently proposed reconfigurable distributed antennas and reflecting surface (RDARS) architecture \cite{ma2023reconfigurable} to realize user localization and enhance communication performance simultaneously.

\section{Technical Design}
\begin{figure}[t]
  \centering
    \includegraphics[width=0.4\textwidth]{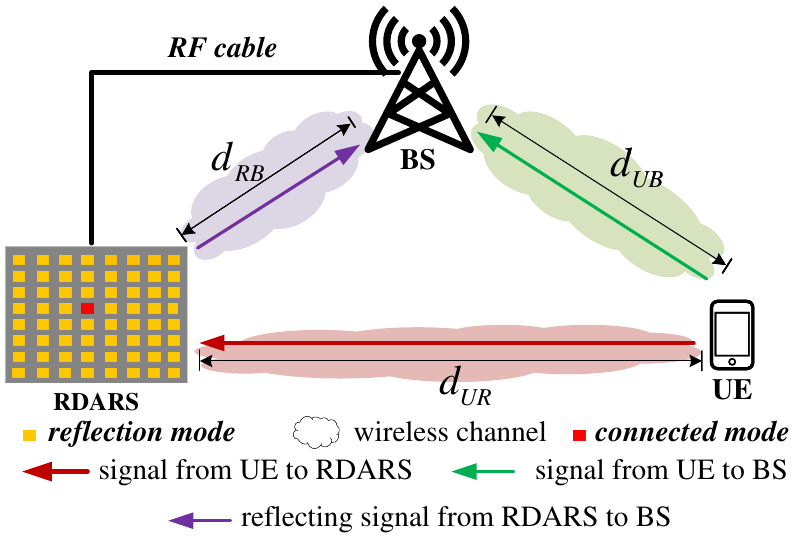}
    \caption{The system architecture diagram.}
    \label{fig: mode}
  \vspace{-0.8cm}
\end{figure}

The architecture diagram of the proposed uplink ISAC system is shown in Fig.~\ref{fig: mode}, where the low-cost and low-energy-consumption RDARS is deployed to assist in the uplink signal transmission and user localization.  

The RDARS \cite{ma2023reconfigurable} is composed of a number of passive elements, each of which can be dynamically programmed to operate on two working modes, namely, the reflection mode and the connected mode as shown in Fig.~\ref{fig: mode}. 
The element under the reflection mode works as the conventional RIS to dynamically adjust the wireless channels into favorable ones.
On the other hand, the element working on the connected mode serves as the remote receive antenna to receive the signal directly. 
Flexible configurations are available for the number and locations of the connected elements. 
A thorough performance analysis and experiment test can be found in \cite{ma2023reconfigurable}. 
By introducing such an appealing structure of RDARS, both the communication and sensing performance will be improved significantly.

\begin{figure}[t]
  \centering    \includegraphics[width=0.4\textwidth]{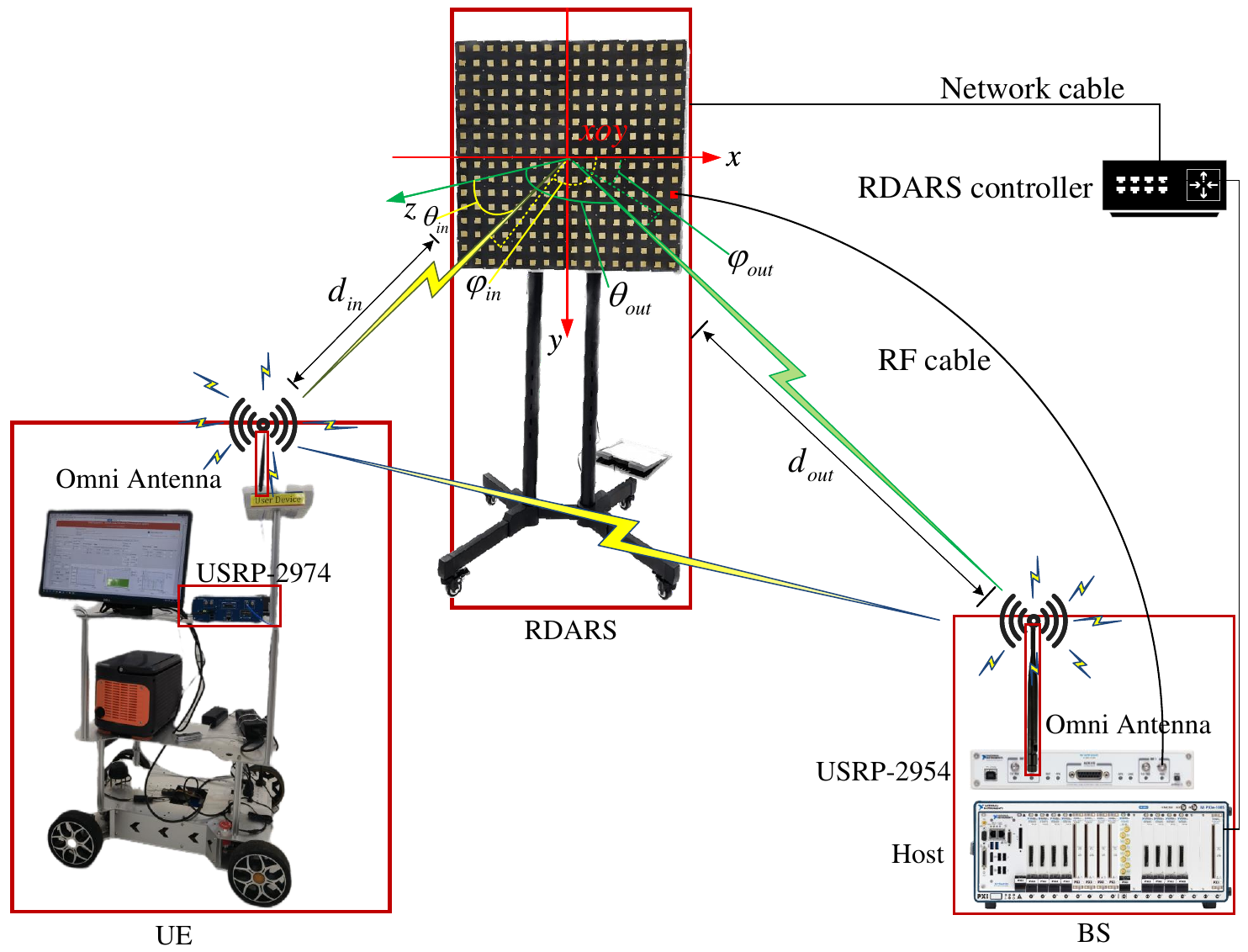}
    \caption{The RDARS-aided ISAC system prototype.}
    \label{fig: architecture}
    \vspace{-0.5cm}
\end{figure}

\subsection{Hardware Description}
 The developed hardware platform of RDARS-aided ISAC system with one base station (BS) and one user equipment (UE), each with a single antenna, is shown in Fig.~\ref{fig: architecture}.
The system is working at 3.7GHz with 20 MHz bandwidth, using a standard 3GPP LTE radio frame structure.
\subsubsection{Transceiver}
 Both the BS and the UE utilize the software-defined radio (SDR) platform, composed of the PXI hardware architecture and the LabVIEW programming software, to realize real-time communication, including video-streaming transmission. 
 Specifically, the BS consists of a USRP-2954, multiple FlexRIO 7976R FPGA modules, and one PXIe-1085 chassis for real-time baseband signal processing, including modulation, channel estimation, fast Fourier transform, etc.
 The UE is equipped with one USRP-2974 for uplink signal transmission. 
 Both signal processing and sensing are performed at the BS. 

\subsubsection{RDARS}
Furthermore, the RDARS operates within the 3.4GHz-3.8GHz frequency band and consists of 256 elements, each equipped with a 2-bit phase shifter when performing reflection mode and also can be switched to the connected mode as needed.  
The mode and the phase shift of each element can be dynamically and independently controlled by the RDARS controller with a total 512-bit control signal delivered through the control link using the User Datagram Protocol (UDP).
In this prototype, we select $a$ elements serving in the connected mode, and the remaining $256\!-\!a$ elements in the reflection mode are encoded to generate a specific reflection beam towards BS. 
Both the number and positions of the $a$ connected elements can be flexibly configured.  



\subsection{Main Algorithm Design}
For the software design, we develop a two-stage sensing scheme to achieve user localization, while maintaining communication performance. 
The two-stage localization scheme includes beam sweeping and range estimation. 

\subsubsection{Beam Sweeping}
Due to the special structure of RDARS,  the 2-bit phase shifter of each element under reflection mode can be programmed to form the designated beam for specific directions. And BS has the capability of controlling the phase of each element working on reflecting mode by generating the unique phase coding matrix.
In view of the beam width of RDARS, the beam sweeping overhead can be reduced by separating the beam space into finite blocks.
With the fixed deployments of BS and RDARS, a 2D azimuth-elevation sweeping scheme is proposed to determine the optimal incident angle based on the measured received signal strength indication (RSSI). 
The proposed beam sweeping scheme can help exploit the angle information of the user without affecting the communication performance.


\subsubsection{Range Estimation}
Based on the obtained optimal azimuth and elevation angles of the user, we first derive the angle between UE-RDARS and BS-RDARS based on the geometry relationship. 
On the other hand, the received power at the BS's antenna and the RDARS's connected elements can be measured based on the RSSI information.  
Even without prior knowledge of path loss parameters, the range between the user and RDARS can be substantially estimated with the measured received powers and geometry equations. 
However, due to the signal fluctuations incurred by shadow fading and imperfect hardware implementation, it's necessary to compensate that by virtue of some calibration algorithms.
Consequently, the location of the user is obtained with the proposed two-stage scheme. 

\begin{figure}[t]
  \centering    \includegraphics[width=0.4\textwidth]{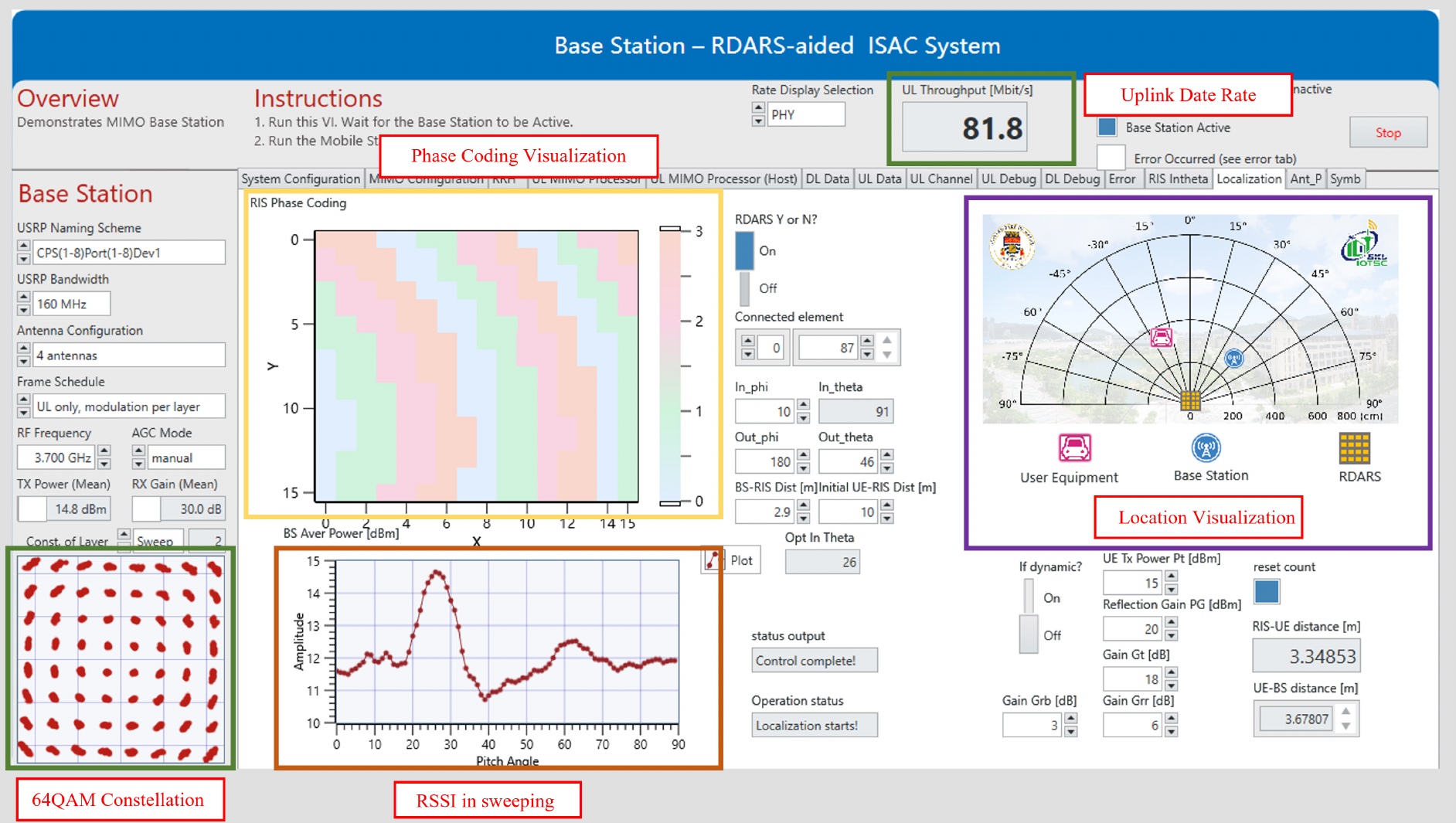}
    \caption{Field test result.}
    \label{fig: test result}
    \vspace{-0.5cm}
\end{figure}
The field test result of the proposed RDARS-aided ISAC system is shown in Fig.~\ref{fig: test result}. 
The RSSI fluctuations are illustrated in the field test during the beam sweeping, with one peak in accordance with the optimal beam.  
And the best output of the beam sweeping is less than 5° from the ground truth. 
The range estimation also shows better performance than some relevant works.
At the same, the data rate of uplink communications with 64-QAM modulation achieves 81.8 Mbit/s, which demonstrates the developed prototype realizes integrated sensing and communication functionalities simultaneously.

\vspace{-2mm}
\section{Conclusion}
In this paper, we develop an uplink RDARS-aided ISAC system prototype to simultaneously achieve communication signal transmission and user location estimation. 
The proposed RDARS-ISAC prototype provides an insightful precedent for integrating the existing communication network with sensing functionalities, such as indoor localization, without any additional hardware resources.

\vspace{-2mm}
\section*{Acknowledgment}
This work was supported in part by the Science and Technology Development Fund, Macau SAR 
under Grants 0087/2022/AFJ and SKL-IOTSC(UM)-2021-2023, in part by the National Natural Science 
Foundation of China under Grant 62261160650.

\vspace{-3mm}
\bibliographystyle{IEEEtran}
\bibliography{RDARS_ISAC_Demo}

\end{document}